\documentclass[aps,reprint]{revtex4-1}
\bibpunct{[}{]}{;}{n}{}{,}
\usepackage{graphicx}
\usepackage{marvosym}
\usepackage{amssymb}
\usepackage{amsmath}

\begin{document}

\title{Spin Localization of a Fermi Polaron in a Quasirandom Optical Lattice}

\author{Callum W. Duncan}
\affiliation{SUPA, Institute of Photonics and Quantum Sciences, Heriot-Watt University, Edinburgh EH14 4AS, United Kingdom}
\author{Niels J. S. Loft}
\affiliation{Department of Physics \& Astronomy, Aarhus University, Ny Munkegade 120, 8000 Aarhus C, Denmark}
\author{Patrik \"Ohberg}
\affiliation{SUPA, Institute of Photonics and Quantum Sciences, Heriot-Watt University, Edinburgh EH14 4AS, United Kingdom}
\author{Nikolaj T. Zinner}
\affiliation{Department of Physics \& Astronomy, Aarhus University, Ny Munkegade 120, 8000 Aarhus C, Denmark}
\author{Manuel Valiente}
\affiliation{SUPA, Institute of Photonics and Quantum Sciences, Heriot-Watt University, Edinburgh EH14 4AS, United Kingdom}

\begin{abstract}
Recently, the topics of many-body localization (MBL) and one-dimensional strongly interacting few-body systems have received a lot of interest. These two topics have been largely developed separately. However, the generality of the latter as far as external potentials are concerned -- including random and quasirandom potentials -- and their shared spatial dimensionality, makes it an interesting way of dealing with MBL in the strongly interacting regime. Utilising tools developed for few-body systems we look to gain insight into the localization properties of the spin in a Fermi gas with strong interactions. We observe a delocalized--localized transition over a range of fillings of a quasirandom lattice. We find this transition to be of a different nature for low and high fillings, due to the diluteness of the system for low fillings. 
\end{abstract}

\maketitle

\section{Introduction}
\label{sec:Intro}

Strongly interacting one-dimensional quantum systems have attracted
major attention in recent years \cite{Wall2013,Volosniev2014,Deuretzbacher2014,Volosniev2015,Hu2016}.
When confined to one dimension the fermionic system exhibits a spin-charge separation, and for very strong interactions the charge degrees of freedom are frozen, making it possible to write an
effective spin chain Hamiltonian for the system \cite{Volosniev2014,Deuretzbacher2014,Volosniev2015}. Methods have
been developed to solve numerically for the exchange coefficients of this spin chain for any given confining potential
\cite{Deuretzbacher2014,Loft2016a}.

The presence of disorder in an interacting system can result in the
violation of the eigenstate thermalization hypothesis, due to
many-body localization (MBL) \cite{Nandkishore2015,Altman2015}. The
localization of single-particle states in the presence of disorder in quantum systems was originally considered by Anderson in 1958
\cite{Anderson1958}. Over the intervening decades, Anderson
localization has been observed in many systems, including in electron gases \cite{Cutler1969}, photonic lattices \cite{Segev2013}, and cold atoms \cite{Billy2008}. For a MBL phase in the tight-binding approximation all eigenstates of the system are Anderson localized \cite{Nandkishore2015}. Theoretical work on MBL has been focused on the nature of the delocalization-localization phase transition as disorder is increased
\cite{Pal2010,Kjall2014,Serbyn2015,Imbrie2016}. Quantum spin chains have been fruitful models for looking at this transition. In most cases the disorder is introduced in the external magnetic field or coupling coefficients of the spin chain. In this work we still consider a quantum spin chain, but one that is induced by the strong interactions present between fermions. We introduce disorder in the system via a quasirandom optical lattice potential.

In recent years, the field of ultracold atomic gases in one dimension has received a lot of interest
\cite{Bloch2005,Bloch2008,Guan2013}. Such systems have been considered for strongly interacting fermions \cite{Pagano2014} and bosons \cite{Kinoshita2004,Paredes2004,Lazarides2011}. In this field, the MBL phase transition has been observed with interacting fermions in a one dimensional quasirandom optical lattice
\cite{Schreiber2015}. Recently, an experimental realization of only a few strongly interacting fermions in a one dimensional trap has been realised \cite{Murmann2015}.

\section{Model}
\label{sec:Model}

\begin{figure}[t]
\begin{center}
\includegraphics[width=0.5\textwidth]{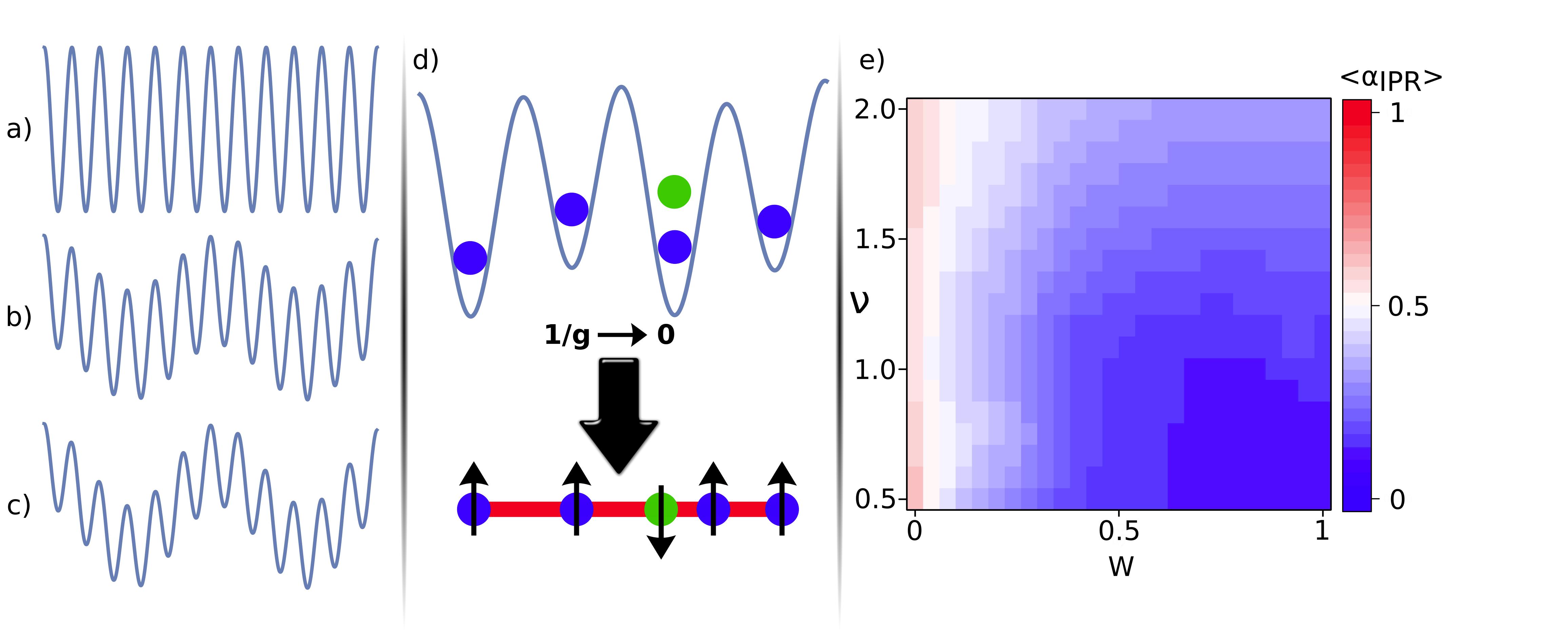}
\end{center}
\caption{a -- c) The quasirandom potential, Eq.~(\ref{eq:Potential}), for $W=0$, $0.5$ and $1$ respectively. d) Illustration of the mapping to an effective spin chain model for strong interactions. e) Average inverse participation ratio for a filling $\nu=N/L_s$ of the $N$ single particle states for disorder strength $W$.}
\label{fig:PotandSingle}
\end{figure}

We consider $N$ strongly interacting repulsive spin-1/2 fermions in one
dimension (see Fig.~\ref{fig:PotandSingle}d). This system is described by the Hamiltonian
\begin{equation}
H = \sum_{i=1}^N \left( \frac{p_i^2}{2m} + V(x_i)\right) + g\sum_{i<j}^N \delta(x_i - x_j)
\label{eq:FullHamiltonian}
\end{equation}
where $g$ is the contact interaction strength, and $V(x_i)$ is the single-particle external potential. We consider the limit of strong repulsive interactions, $g \rightarrow +\infty$, for which the system can be mapped onto an effective spin chain model. We will elaborate on this below. Throughout this paper we set $\hbar = m = 1$, and express length in units of the length of the system $L$.

Quasirandom, or quasiperiodic potentials have been shown to exhibit a localization transition for single particle \cite{Grempel1982,Ganeshan2015}, and many-body systems \cite{Iyer2013}, as is the case for truly random disorder. Such potentials can be implemented in ultracold atom set-ups \cite{Fallani2008}, and have been used to observe both Anderson localization \cite{Roati2008}, and MBL \cite{Schreiber2015}. We
consider a quasirandom potential with open boundary conditions, with a main lattice of strength $V_1$ and a disorder term of strength $V_2$. The potential $V(x)$, appearing in Eq.~(\ref{eq:FullHamiltonian}) is given by
\begin{equation}
V(x) = V_1 \cos\left(\frac{\tau_1 x}{d}\right) + V_2 \cos\left(\frac{\tau_2 x}{d} + \phi\right),
\label{eq:Potential}
\end{equation}
where $d$ is the lattice spacing, defined as $d \equiv L/L_s$ with $L_s$ giving the number of wells -- or `sites' -- in the lattice. Throughout this work we set $\tau_1 = 2 \pi$ and $\tau_2 = 1$, satisfying the need for $\tau_1 / \tau_2$ to be incommensurate for the above potential to be quasirandom. We fix the number of lattice wells $L_s = 12$, and sweep across the lattice filling ($\nu \equiv N /L_s$) by varying the number of particles $N = 6,7,8,\dots,24$. We will quantify the disorder strength by the ratio $W = V_2 / V_1$, and consider the disorder range of $0 \leq W \leq 1$, with examples of the potential shown in Fig.~\ref{fig:PotandSingle}a, b and c. The main lattice strength $V_1 = 5$ is chosen to ensure that the lattice is strong enough to be felt by all particles, without the particles being localized into single sites.

\begin{figure*}[t]
\begin{center}
\includegraphics[width=0.6\textwidth]{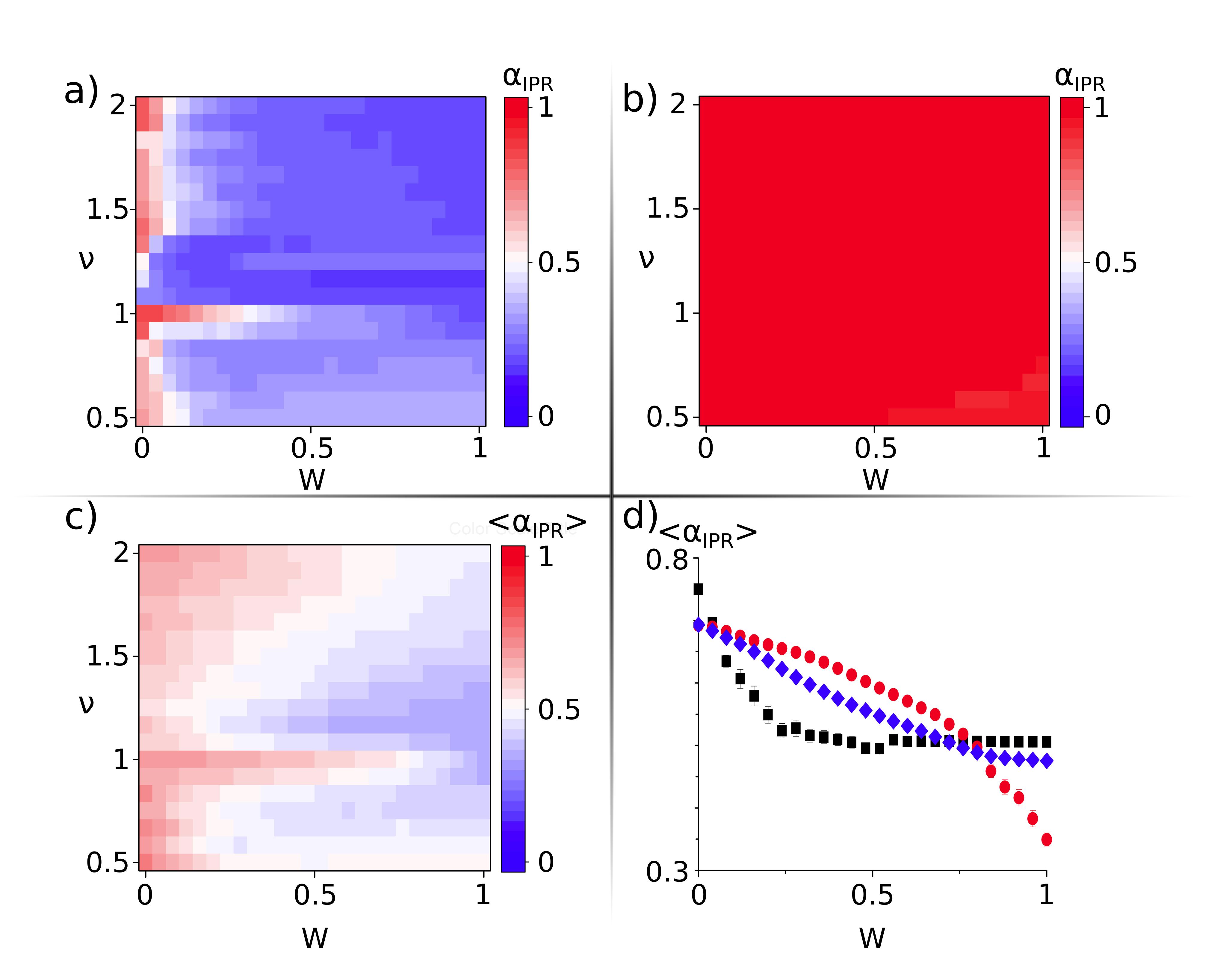}
\end{center}
\caption{$\alpha_{\text{IPR}}$ for the single polaron spin chain. a) Groundstate IPR.  b) highest energy state IPR. c) The average IPR over all states except the ground and highest, $\langle \alpha_{\text{IPR}} \rangle$. d) cut-outs of c) for $\nu = 0.5$  squares (black), $\nu = 1$ circles (red), and $\nu = 2$ diamonds (blue).}
\label{fig:SinglePolaron}
\end{figure*}

In the case of strong repulsion, $g \rightarrow \infty$, the system of trapped cold atoms can be described by an effective spin chain model \cite{Volosniev2014,Deuretzbacher2014,Volosniev2015,Loft2016a}. Specifically, to linear order in $1/g \ll 1$ the (ground state manifold) spectrum is given by
\begin{equation}
E_n = E_0 - \frac{K_n}{g} \; ,
\end{equation} 
with $E_0$ being the degenerate many-body ground state energy at infinitely strong repulsion $1/g=0$. Here $K_n$ are the eigenstates of the spin chain Hamiltonian  
\begin{equation}
K = - \frac{1}{2} \sum_{j=1}^{N-1} J_j (\boldsymbol{\sigma}_j \cdot \boldsymbol{\sigma}_{j+1} - 1),
\label{eq:SpinHamiltonian}
\end{equation} 
with $\boldsymbol{\sigma}_j$ being the Pauli matrices acting on the $j$th site (or atom) of the spin chain, and $J_j$ is the coefficient connecting the $j$ and $j+1$ sites. The spin chain coefficients $J_j$ are solely dependent on the single-particle wavefunctions, which are found by solving the stationary Schr\"odinger equation with the single-particle Hamiltonian $H_0 = p^2/2m + V(x)$. Thus different realizations of the quasirandom potential $V(x)$ will translate into variations in the spin chain coefficients for the effective spin chain~(\ref{eq:SpinHamiltonian}). We use the open source program CONAN \cite{Loft2016a}, which has been developed to take arbitrary potentials and numerically calculate the $N-1$ coefficients $J_j$ between the spin chain sites for up to $N \sim 30$ particles. Notice that in this approach, we study the spin chain model resulting from \emph{every single} realization of the quasirandom potential. The above spin chain model is a pertubative description that is exact to linear order -- therefore variational -- in $1/g$ of the ground state manifold, with two assumptions: Firstly strong repulsion, and secondly zero-temperature. Recently, the formation of an effective spin chain in such limits has been confirmed with agreement between numerics and an experimental system using only a few cold atoms in a one-dimensional harmonic trap \cite{Murmann2015}.

For the numerical investigations we compute the spin chain coefficients $J_j$, using CONAN \cite{Loft2016a}, arising from the lattice potential in Eq.~\eqref{eq:Potential} for $W$ between 0 and 1, and over a range of particle numbers $N = 6,7,8, \dots ,24$, corresponding to fillings $\nu \equiv N/L_s = 1/2,0.583,0.667,\dots,2$. For each $W$ and $N$ we average over 19 realizations of the phase $\phi$. Using the calculated spin chain coefficients, we solve the stationary Schr\"{o}dinger equation for the spin chain Hamiltonian,
Eq.~\eqref{eq:SpinHamiltonian}, numerically. For the polaron we will denote the wavefunction as
\begin{equation}
\mid \! \Psi \rangle = \sum_{j=1}^N C_j \mid \uparrow \dots \uparrow \left(\downarrow\right)_j \uparrow \dots \uparrow \rangle,
\end{equation}
where $C_j$ is the coefficient for the polaron in the $j$th spin chain site. To gain further insight, we will also consider the case of two polarons, which we expect to have similar general behaviour to the single polaron in this system. For two polarons we write the
wavefunction as
\begin{equation}
\mid \! \Psi \rangle = \sum_{i < j}^N C_{(i,j)} \mid \uparrow \dots \uparrow \left(\downarrow\right)_i \uparrow \dots \uparrow \left(\downarrow\right)_j \uparrow \dots \uparrow \rangle.
\end{equation}

\section{Measures of Localization}
\label{sec:Measures}

The onset of Anderson localization in the system can be observed by considering the inverse participation ratio (IPR) \cite{Hu2016}, given by
\begin{equation}
\alpha_{\text{IPR}} = \frac{1}{m \sum_{v} \mid C_{v} \mid^4},
\label{eq:IPR}
\end{equation}
with $C_{v}$ being the coefficients of either the polaron $v=j$ or two polaron $v=(i,j)$ states, and $m$ denoting the size of the Hilbert space of the wavefunction. For a fully delocalized state $\alpha_{\text{IPR}} \sim 1$. For a fully localized state we have a convergence of $\alpha_{\text{IPR}}$ towards zero. We will consider the IPR of the ground, and the highest energy states. In addition we calculate the average IPR across all other states (denoted by $\langle \alpha_{\text{IPR}}\rangle$). The average IPR gives an indication of the overall localization of the system. However, this is not an exact measure of the localization of all states, e.g. there could be a few heavily localized states with the rest delocalized.

A standard measure of the MBL transition is the properties of the energy level statistics, which can be investigated via the ratio of adjacent energy level gaps \cite{Hu2016,Serbyn2015,Oganesyan2007,Mondaini2015}
\begin{equation}
r_n = \frac{\mathrm{min}(\delta_n,\delta_{n-1})}{\mathrm{max}(\delta_n,\delta_{n-1})},
\label{eq:GapRatio}
\end{equation} 
where $\delta_n = E_n-E_{n-1}$ is the gap in the spectrum between the $E_n$ and $E_{n-1}$ eigenvalues, with $\mathrm{min}$ and $\mathrm{max}$ taking the minimum and maximum value of the two adjacent gaps in the spectum $(\delta_n,\delta_{n-1})$, ensuring $0 \leq r_n \leq 1$. We will take the average of the gap ratio $\langle r_n \rangle$ over all $\delta_n$, with the exclusion of the ground and highest energy states as will be discussed in Sec.~\ref{sec:Results}. In the delocalized phase we expect the energy level statistics to satisfy a Wigner-Dyson disitribution (WD) \cite{Serbyn2015,Oganesyan2007}, with the average ratio $\langle r_n \rangle_{\text{WD}} \simeq 0.536$ \cite{Hu2016}. Meanwhile, the MBL phase has statistics that satisfy the Poisson distribution (PD) \cite{Serbyn2015,Oganesyan2007}, with an average ratio of $\langle r_n \rangle_{\text{PD}} \simeq 0.386$ \cite{Hu2016}.

\section{Localization of the Spin}
\label{sec:Results}

\begin{figure*}[t]
\begin{center}
\includegraphics[width=0.75\textwidth]{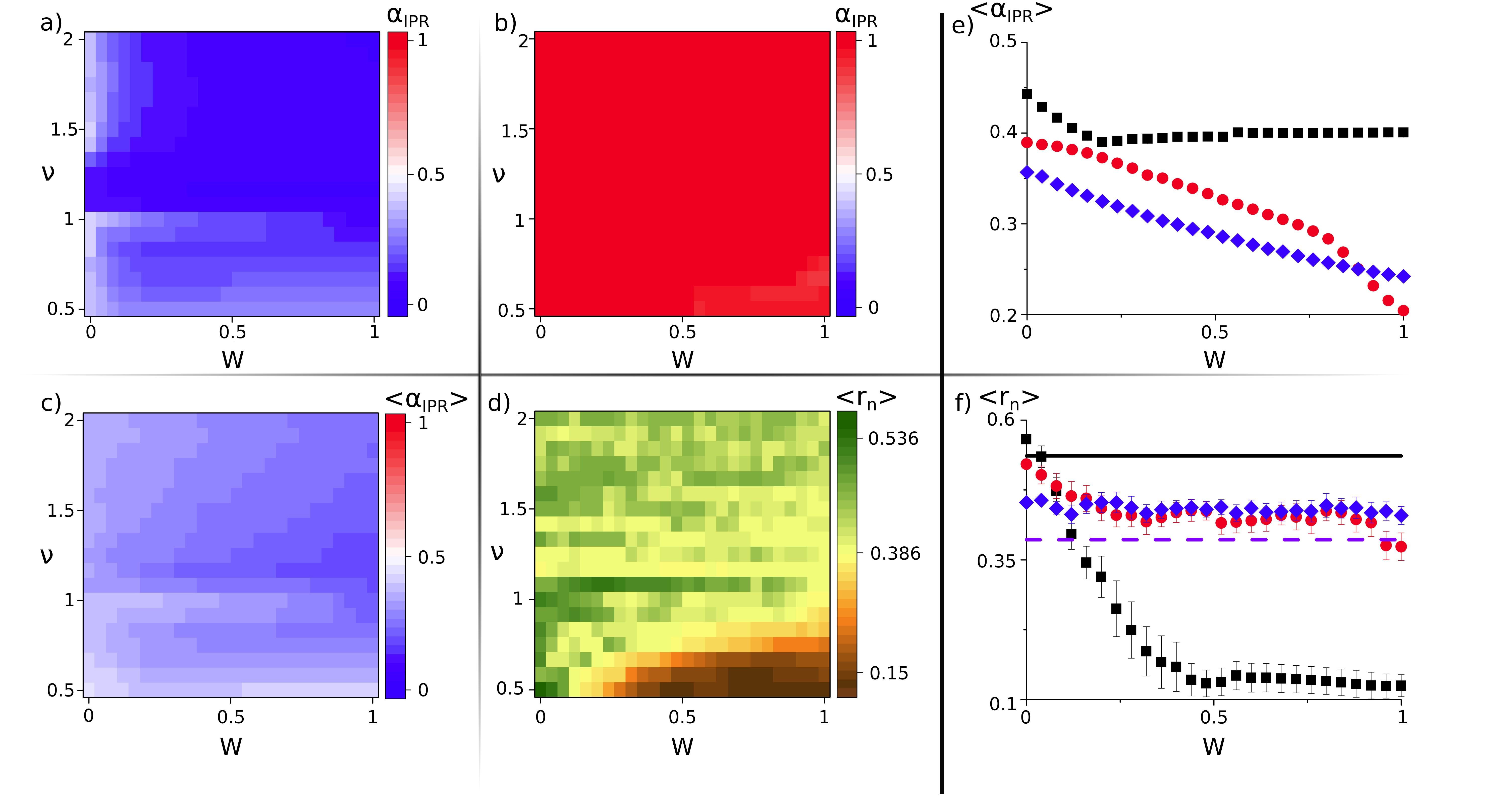}
\end{center}
\caption{$\alpha_{\text{IPR}}$ for the two-polaron spin chain. a) Groundstate IPR, averaged over 9 disorder phase, $\phi$, realizations.  b) Same as a) but for the highest energy state. c) The average IPR over all states except the ground and highest, $\langle \alpha_{\text{IPR}} \rangle$. d) The energy gap ratio, $\langle r_n \rangle$. e) and f) cut-outs of c) and d) respectively for $\nu = 0.5$  squares (black), $\nu = 1$ circles (red), and $\nu = 2$ diamonds (blue). On plot f) black solid line and purple dashed line denote the expected values for an extended and localized phase respectively.}
\label{fig:TwoPolaron}
\end{figure*}

First we confirm the localization in the charge degree of freedom. In Fig.~\ref{fig:PotandSingle}e, we consider the average IPR over all single particle states for the quasirandom lattice, Eq.~(\ref{eq:Potential}). We find for disorder above $W \sim 0.5$ the general localization of the single particle states across the whole range of $\nu$. The critical disorder for the delocalized to localized transition of the single particle states is weakly dependent on the filling of the lattice. This result gives a good indication that the particles are ``feeling" the lattice potential.

For the spin of a single polaron we consider $\alpha_{\text{IPR}}$ of the ground, and highest energy states, then $\langle \alpha_{\text{IPR}}\rangle$ over all other states, with the results shown in Fig.~\ref{fig:SinglePolaron}. In addition, we consider two polarons in the system, with the same observables as for the single polaron, but in addition we calculate the average energy gap ratio $\langle r_n \rangle$. We do not consider the gap ratio for the single polaron due to the small number of states, $N$, in the system, resulting in large variances in $\langle r_n \rangle$ over the realizations of the disorder. Naturally, for two polarons there is a larger number of states, $N(N-1)/2$, resulting in smaller variance over the disorder realizations.

The groundstate of the spin is found to localize at small disorder in Figs. \ref{fig:SinglePolaron}a and \ref{fig:TwoPolaron}a, with strong localization for $W > 0.1$, for most $\nu$. With the exception of around unit filling, where we have a spin in each lattice site, resulting in an elongated transition to the localized state. The highest energy state is delocalized across the system over all disorder, Figs. \ref{fig:SinglePolaron}b and \ref{fig:TwoPolaron}b. Therefore, for our system we can never have a true MBL phase, in the sense that all states will not localize. However, with the inherently delocalized highest energy state excluded, we observe a delocalized-localized transition over a range of $\nu$.

With the average IPR over all states in the system except the ground and highest states, $\langle \alpha_{\text{IPR}} \rangle$, we can gain an insight into the general localization properties of the system, see Figs. \ref{fig:SinglePolaron}c and \ref{fig:TwoPolaron}c. We observe a defined transition from a majority of states being delocalized to heavily localized over a range of fillings from $ 1 \leq \nu \leq 2$. For small fillings, $\nu < 1$, we observe a trend towards localization with large disorder. The relatively weak localization of states at these fillings is due to the diluteness of the system. Each fermion (or groups of fermions) can be well seperated from its neighbours, resulting in weak coupling coefficients, effectively resulting in the separation of the spin chain into sections. Hence we observe some localization of the state, but not due to disorder in the spin chain. The regimes discussed are well shown by the cut outs of Figs. \ref{fig:SinglePolaron}d and \ref{fig:TwoPolaron}e.

However, the IPR is a poor measure of the localization of all states, and a standard measure for this (the MBL phase) is the average energy gap ratio in the spectrum, $\langle r_n \rangle$. We calculated $\langle r_n \rangle$ for two polarons in the system, Fig.~\ref{fig:TwoPolaron}d, with a cut out at select fillings in Fig.~\ref{fig:TwoPolaron}f. For $\nu \sim 1$, we see a transition from an extended ($\langle r_n \rangle_{\text{WD}} \simeq 0.536$) to a localized phase ($\langle r_n \rangle_{\text{PD}} \simeq 0.386$). At $\nu = 1.0833 = 1 + 1/12$, where we are at one particle over unit filling, we observe the states to have Poisson statistics independent of disorder, shown by the yellow region above unit filling in Fig.~\ref{fig:TwoPolaron}d. This is due to the spin chain coefficients having a form that is `well-like' at this filling without the prescence of disorder \cite{Duncan2016}. Thus the statistics of the eigenvalue gaps are that of the Poisson distribution, as has been shown for interacting trapped bosons in harmonic potentials \cite{Chakrabarti2012}. For higher filling, we see a transition from a delocalized to a localized phase with increasing disorder. However as we approach $\nu = 2$, $\langle r_n \rangle$ is consistently at an intermediate value, Fig.~\ref{fig:TwoPolaron}f, which is consistent with a mixed phase of localized and delocalized states.

With $\nu < 1$, we observe a different regime of the system. $\langle r_n \rangle$ converges to a value well bellow $0.386$, as seen in Figs.~\ref{fig:TwoPolaron}d and f, with a weak localization across all states as seen in Figs.~\ref{fig:SinglePolaron}c and \ref{fig:TwoPolaron}c. The convergence value of $\langle r_n \rangle$ is not consistent with any spectrum we know of. For $\nu < 1$ the states are localized because of the break up of the spin chain due to the diluteness of the system, and not through disorder. The gap ratio further reflects the different nature of the localization transition of the states for low filling.

\section{Conclusions}

Using recent advances in describing strongly interacting confined particles in one dimension, we have investigated the localization properties of the spin degree of freedom. It is well known that the charge degree of freedom is localized in this system in the presence of strong interactions. For the spin we observe the localization of the majority of states upon sufficient disorder for $\nu > 1$. For small fillings, $\nu < 1$, we observe a weak localization regime due to the system being dilute. The system considered can never be completely localized, due to the presence of a fully delocalized highest energy state. This state is an inherent property of the system. However with the exclusion of this delocalized state, we observe a delocalized to localized transition for both the polaron and two polaron systems. This transition is seen for fillings above unity by the convergence of the level statistics to the Poisson distribution expected in the MBL phase. For low fillings and above a certain disorder strength we see the emergence of a regime with different statistics, due to the diluteness of the system.

\section*{acknowledgement}

C.W.D. acknowledges support from EPSRC CM-CDT Grant No. EP/L015110/1. P.\"O. and M.V. acknowledge support from EPSRC EP/M024636/1. N. J. S. L. and N. T. Z. acknowledge support by the Danish Council for Independent Research DFF Natural Sciences and the DFF Sapere Aude program.

\end{document}